\documentclass[aps,pre,showpacs,twocolumn]{revtex4}
\usepackage{graphicx,epsfig}
\usepackage{textcomp}
\usepackage{gensymb}

\begin{document}
\title{Interaction of multiple spiral rotors in a reaction-diffusion system}

\author
{Hrishikesh Kalita and Sumana Dutta{\footnote{sumana@iitg.ac.in}}}

\affiliation{Department of Chemistry, Indian Institute of Technology Guwahati,\\
 Guwahati 781039, INDIA. }

\begin{abstract}
 Rotors of reaction and diffusion, that revolve around a circular singularity, and the corresponding spiral wave activity around it, influence the nature of several excitable media. Their dynamics are known to be affected by target waves and external gradients. When the core of two spirals come very close to each other, they could either repel or attract, depending on their relative chirality and the distance separating them. Multiple pairs of spiral waves, in close vicinity, could alter the paths of the individual rotors. A single spiral core, will be influenced most by the rotor closest to it, or linked to it by a common wave. If within a limiting distance, they would attract each other, and annihilate. Else, they will push each other away, until they reach a critical distance, beyond which, all interactions seem to cease. We have carried out numerical simulations based on a reaction-diffusion model to show the possible interactions of multiple spiral rotors. We are able to find a quantitative measurement of the distances determining the kind of interaction. Some unique dynamics is also unraveled. Experiments with the Belousov-Zhabotinsky reaction have successfully demonstrated the validity of our numerical results.
\end{abstract}

\pacs{05.45.-a, 82.40.Ck, 82.40.Qt}

\maketitle

\section{Introduction}

Two-dimensional spiral waves and their three-dimensional counterparts, the scroll waves, are responsible for arrythmia occurring in cardiac systems. The presence of such waves are harbingers of fibrillation in the atria and ventricles of the heart, oftentimes leading to fatal cardiac arrest  \cite{jalifebook, has}. In systems like the social amoeba, the Dictyostelium Discoideum, such spiral structures enable the colony to feed on limited food resources, and aptly interact with each other \cite{dict, aran2}. They are also found in other systems that spans across physics, chemistry, biology and geology \cite{epst,ertl,xenopus}. Hence, the interaction and control of such spiral rotors is of interest to scientists across disciplines.

In the current manuscript we are mainly interested in exploring the dynamics of spiral rotors around each other. It is known that spiral waves can be controlled by external gradients and target waves. Light \cite{mull,aglad} and electric field \cite{ost} has been used to move the tips of the spirals in a two-dimensional reaction-diffusion system, sometimes leading to annihilation of the waves. High-frequency wave trains have been successfully used to force spirals into defect-drifts \cite{sdost}. Using multiple wave-fields, several rotors could even be localized to a particular position. Numerical studies on the FitzHugh-Nagumo model showed that both like as well as unlike charged spiral vortices (topological charges) forms a bound pair possessing either an axis or a center of symmetry \cite{erma}. This charge is attributed to the chirality or the sense of rotation of spiral waves. Co-rotating spirals are considered to have like charges, and counter-rotating spirals are said to be oppositely charged. In yet another study with the Complex Ginzburg Landau equations, it was established that the interaction between well-separated spirals in the particular system is exponentially weak and does not depend on the topological charges \cite{aran}. Symmetry breaking instabilities in a bound state of a spiral pair can also induce one spiral collapse \cite{yama}. However, such studies on the interaction of spiral rotors is mainly limited to a pair of spirals.

Experimentally, instances of suppression and expulsion of one spiral tip by another (one spiral collapse) was also observed in the aggregation of Dictyostelium amoeba \cite{aran2}. This study  demonstrated that a spiral competition instability was responsible for the symmetry breaking of the spiral pair. Numerical studies on model systems were also expanded to multi-armed vortices\cite{pertsov}. Weijer \textit{et. al.}, established from extensive simulations that spirals having same chirality, with tips less than one wavelength apart, form multi-armed spirals \cite{weijer}.

In three dimensions, experiments with co-planar scroll rings demonstrated that their filaments undergo crossover collision and reconnect when they are within a core-length of each other \cite{sdrecon, sdrecon2}. On the other hand, the filaments repelled, when placed over one another. In yet another study of straight, parallel scroll waves, it was established that the filaments repelled only when the inter-filament distance was shorter than the wavelength of the scroll waves \cite{marcus3D}. When this distance was almost equal to the wavelength, the two scroll waves synchronized. The detailed study on the interaction of multiple spiral vortices with identical frequencies, showing spontaneous annihilation and repulsion, and also establishing the exact distances at which the nature of the interaction changes, is yet to be carried out. It remains to be seen if the binding distances for the two dimensional spiral waves are exactly same as the three dimensional scroll waves, or they do differ.

In the current investigation, we show with experimental evidence, spontaneous annihilation of vortices (without any external force), in a media with multiple spiral rotors. We carry out detailed simulations with the Barkley model, for pairs of one, two, and four spirals, by varying the mutual distance between them. It is observed that, with increasing distance between the rotors, attractive potentials become repulsive. Our simulations reveal different interaction zones with a transition from an attractive to a repulsive zone and a further transition from this repulsive zone to a zone of no-interaction. The important role of the wave connecting two rotors has also been established. We also carried out experiments in thin layers of a chemical reaction-diffusion system. We have chosen the Ferroin-catalyzed Belousov-Zhabotinsky(BZ) \cite{epst} reaction for the study of the spiral dynamics. Several experiments have been carried out, varying the distances between pairs of spiral tips. Our experimental results corroborate well with the numerical predictions. The phenomena of spiral repulsion and spiral attraction leading to annihilation, have been successfully demonstrated, for a system of upto eight rotors.

\section{Numerical model}
The generic two-variable Barkley model is often employed for the study of reaction-diffusion systems \cite{bark2}. It has been widely used for the study of spiral and scroll waves in the BZ system \cite{ost,barkbz1,sd2}.
In the presence of diffusion, it can be written as
\begin{eqnarray}
\frac {du} {dt} &=&  \frac{1}{\epsilon} \left[ u\left(1-u \right) \left(u-\frac {v+b}{a}\right)\right] +  D_u \nabla^2 u,\label{eq:31}\\
 \frac {dv} {dt} &=& \left(u-v \right) +  D_v \nabla^2 v.\label{eq:32}
\end{eqnarray}
Here, $u$ is the activator and $v$ the inhibitor. In the BZ-system, $u$ and $v$ are loosely related to the concentrations of bromous acid and the oxidized form of ferroin, the catalyst, respectively. $D_u$ = $D_v$ = 1.0, are the diffusion coefficients of the two species. For our simulations, we have chosen the parameter values of $a = 0.84$, $b = 0.07$ and $\epsilon= 0.02$.

The fourth order Runge Kutta method was used to integrate the differential equations. A proper Laplace stencil and appropriate initial conditions were employed to solve the equations. The system was discretized into a $300 \times 300$ lattice. We employed no flux boundary conditions on all sides. A time interval of $\Delta t = 0.012$ time units and a step size of $\Delta x= 0.35$ space units was chosen. This choice of parameter values can initiate and sustain stable, non-meandering spirals, with a circular core of diameter $d_s=1.8$ space units. The average wavelength of the spirals ($\lambda$), far from the core is around 18.2 space units.

\begin{figure}[t!]
\centering
\includegraphics[width=0.5\textwidth]{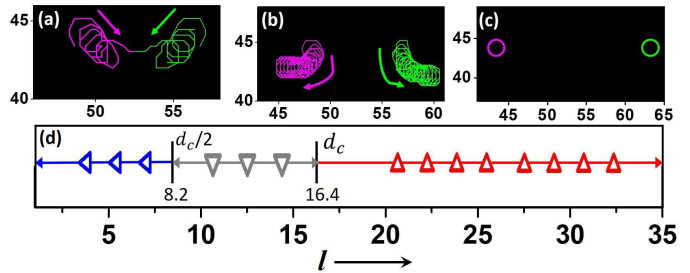}
\caption{Interaction between the two cores of a spiral-pair. (a) Attraction leading to annihilation for $l= 6.65$ space units, (b) repulsion between the spiral rotors, for $l=9.25$, (c). No visible interaction between the rotors for $l=19.75$. (d) Phase diagram of interaction between the two cores in a spiral pair. The open triangles depict the type of interaction for a given simulation with fixed $l$ value. Blue left-pointing triangles portray horizontal annihilation [like in (a)], grey down-ward pointing triangles depict repulsion [like in (b)], and the open red up-ward pointing triangles represent the absence of any interaction between the two rotors [as in (c)]. $d_c$ depicts the critical distance of interaction.}
\label{fig1}
\end{figure}

\section{Numerical results and Discussions}

We begin our simulations with the simplest case: a single spiral pair. When the distance between the center of the two rotors ($l$) is quite large, e.g. $l=19.75$ space units in Fig. \ref{fig1}(c), the two spirals keep on rotating around their circular cores. As the cores are brought closer, we can observe a repulsion between them, which pushes them apart, till the distance between them is a little less than one wavelength. Figure \ref{fig1}(b) shows one such experiment, where initial value of $l$ is 9.25. However, in this case, the right tip travels a little lower than the left one, breaking the symmetry slightly. As we keep decreasing the initial distance between the cores, this repulsion phenomenon is observed until $l=8.2$, below which there is a sudden change from repulsive behavior to an attractive one. In Fig. \ref{fig1}(a), an example of attractive interaction between the cores of a spiral pair, which are initially 6.65 space units apart, is seen. The two vortices are observed to trace a curved path that brings them close together, and finally annihilate. The annihilation of two counter-rotating spirals had also been previously observed in literature, when they lie within one core-length of each other \cite{erma}. However, in this case, we find that such spirals attract even when they are further away, as long as they are less than 8.2 space units apart.  When the distance between the spiral tips is less than 1.8 space units, the spiral rotors do not complete one full rotation. Instead, their strong mutual attraction forces them to annihilate before that. It is to be noted that 1.8 space units is the diameter of a non-interacting spiral core for the chosen parameter range.

A phase diagram constructed on the behavior of two interacting spiral cores is shown in Fig. \ref{fig1}(d). There are three clearly marked regions in the phase diagram, one of attraction leading to annihilation ($8.2>l$), repulsion ($16.4 \geq l \geq 8.2$) and no-interaction ($l> 16.4$). Interestingly, we may observe that when $l > \lambda - d_s =16.4$, the spirals do not interact. Let us call this value the critical distance, $d_c$. On the other hand, the interaction changes from repulsive to attractive at exactly half of this critical value, i.e. $l = \frac{1}{2}(\lambda - d_s)=8.2$. For our simulations, the spiral tips begin to rotate at a maximum distance ($l_{max}=l + d_s$), and then come closer to each other as the two tips rotate in. The minimum distance between the tips, for a given $l$ value is $l_{min}=l-d_s$. This means, that the maximum distance $l_{max}$ between the two spiral tips must be more than one wavelength ($\lambda$), for there to be no interaction between them.

\begin{figure}[t!]
\centering
\includegraphics[width=0.45\textwidth]{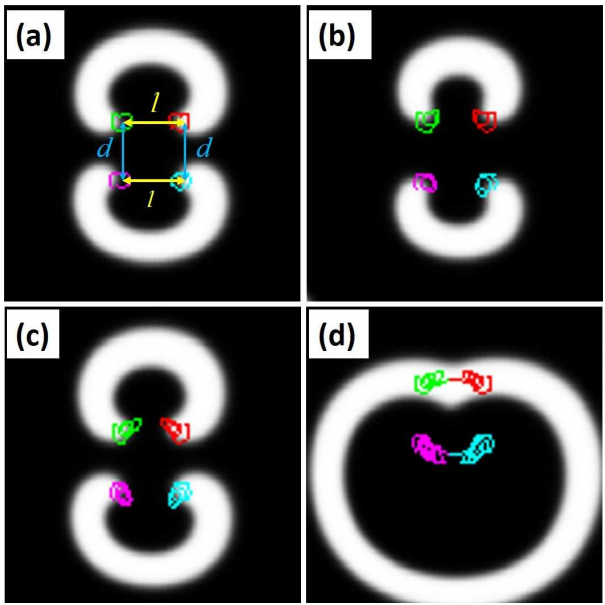}
\caption{Interaction of two spiral pairs showing attraction and annihilation. Snapshots of spirals along with tip-trajectories at $t$ = (a) 7.2, (b) 12.0, (c) 18.0, (d) 31.2, in normalized time units. The colored curves are the tip-trajectories. Each tip has been assigned a unique color for better visualization. The initial horizontal inter-tip distance $l= 7.4$ (yellow horizontal straight lines), and initial vertical distance between cores $d=7.0$ (blue vertical straight lines), are marked in (b). Area of each snapshot is 35 space units $\times$ 35 space units.}
\label{fig2}
\end{figure}

Next, we carry out simulations with two spiral pairs (a four rotor system). We made the system completely symmetrical by keeping the distance($l$) between the two cores in each spiral-pair, the same. We also maintain equal distance between the two spiral pairs ($d$), for a particular simulation [please refer Fig. \ref{fig2} (a)]. We carried out several simulations by varying the horizontal distance between the cores ($l$) and the vertical nearest neighbor distance ($d$). By doing so, we observed different kinds of interactive phenomenon between the spiral tips.

\begin{figure}
\centering
\includegraphics[width=0.45\textwidth]{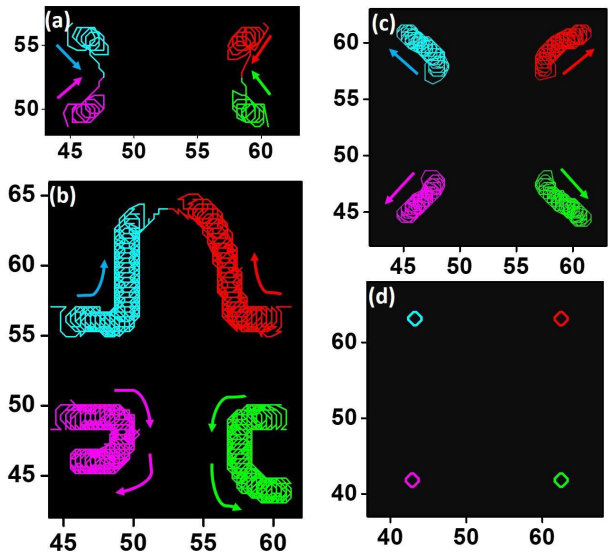}
\caption{Interaction between pairs of spirals. Tip trajectories showing (a) Vertical annihilation for both pairs, where initially,  $l=15.75$ and $d=6.3$. (b) Simultaneous repulsion between one pair and annihilation between the other, for  $l=14.35$ and $d=7.0$. (c) Repulsion between all rotors when  $l=10.15$ and $d=9.8$. (d) No interaction for  $l=19.3$ and $d=21.2$.}
\label{fig3}
\end{figure}

Figure \ref{fig2} depicts an example, where mutual collapse occurs between the cores of a spiral-pair, due to annihilation along $l$. We later refer to this kind of annihilation as horizontal annihilation. The tip trajectories highlight the attraction of the vortices. As time progresses, the tips get closer, indicating an attractive interaction between the two cores of a spiral-pair. In this example, the attraction takes place sidewise amongst the two spiral rotors of a pair. Here $l=7.4$ and $d=7.0$, both smaller than the critical distance, $d_c=8.2$. However, that $l$ is larger than $d$ has no effect on the horizontal attraction of the spirals. It indicates a stronger attraction between the pair of spirals that are initially linked by the same wave, compared to a pair that belong to neighboring waves.

\begin{figure}[t!]
\centering
\includegraphics[width=0.5\textwidth]{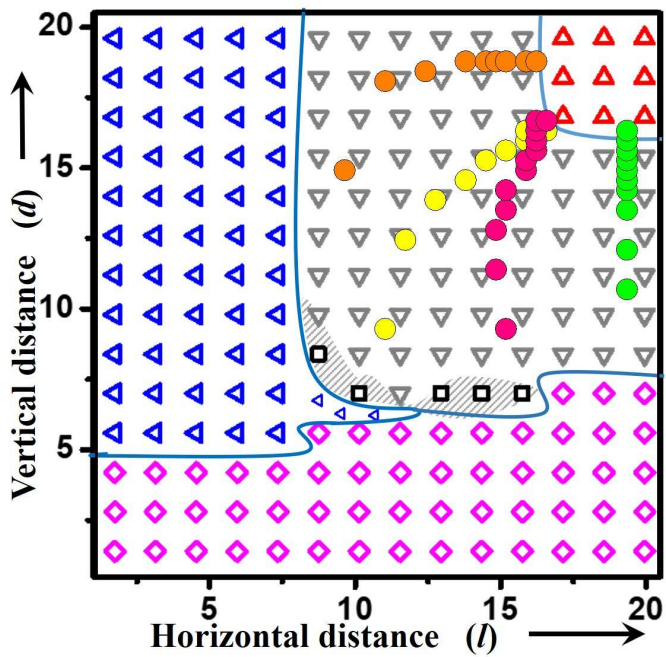}
\caption{Phase diagram of interaction between two pairs of spirals. The open triangles, diamonds and squares depict the type of interaction for a given simulation with fixed $l$ and $d$ value. Magenta diamonds depict vertical annihilation, the blue left-pointing triangles depict horizontal annihilation, grey down-ward pointing triangles depict repulsion, and the open red up-ward pointing triangles depict no-interaction between any rotor. The black squares are the cases where one of the pairs attract and annihilate, while the other pair repels and moves away from each other. The various regions have been separated by solid lines, which are only qualitative in nature. The shaded region denotes the occurrence of the complicated dynamics of spiral tips as in Fig. \ref{fig3}(b). Four numerical experiments have been traced here, with the changing $l$ and $d$ value of their spiral cores depicted by closed colored circles of a unique color. It shows how the rotors move from different zones of repulsive interaction, into the non-interacting zone.}
\label{fig4}
\end{figure}

Figure \ref{fig3} shows all other kinds of possible interactions between spiral pairs, as the $l$ and $d$ values are varied. Figure \ref{fig3}(a) is an instance of annihilation of vortices, between two spiral pairs facing each other. Two spiral tips that initially belong to different spiral pairs, start attracting each other, along the vertical distance $d$, until they finally annihilate each other. This event is later referred to as horizontal annihilation. The spiral rotors might have chosen a horizontal annihilation over the vertical in this example, because of the very low value of $d$(=6.3) as compared to $l$(=15.75), the latter 2.5 times the value of $d$.

Figure \ref{fig3}(b) illustrates some remarkable phenomena. Here $l=14.35$ and $d=7.0$. One would expect the spirals to annihilate vertically. Instead, there is an initial attraction along the length $l$, for both spiral pairs. After covering a certain distance however, the spiral rotors become strongly repulsive, in the direction of $d$.  For the bottom pair, the spiral tips approach each other initially, and then make a complete U-turn, and go apart horizontally. On the other hand, in the top pair, they attract for a while, and then start moving parallelly away from the other pair, before once again attracting each other, finally leading to the annihilation of the spiral pair. This is a very good example of a symmetry breaking phenomenon. It also indicates a cooperative effect among the spirals. In Fig. \ref{fig3}(c), initially, the repulsion is along $d$ but after some time it starts acting in both the directions making the tips trace almost diagonal trajectories away from their initial positions. With very high $d$ and $l$ values, the dynamics of the waves become independent of each other (no interaction), as in Fig. \ref{fig3}(d).

Figure \ref{fig4} summarizes the results of all our numerical experiments with two spiral pairs. This phase diagram gives us a better understanding of the interactive effects and their dependance on the distances, $l$ and $d$. When the distance between two rotors is very low, they show strong attraction towards each other leading to mutual annihilation. Depending on the values of $l$ and $d$, they either attract horizontally or vertically. Higher limiting values of $l$ compared to $d$ for annihilation indicates a strong attractive power between spirals that are initially joined with the same wave. With increasing $l$ and $d$ values, there is a transition from attractive zone to the repulsive zone. The transition can not be defined by a sharp ($l$, $d$) line, instead it is a thin region, within a range of $l$ and $d$ values. The systems in the transition region could have three kinds of interactions: the spirals may attract and annihilate, they might all repel, or there could be a case of simultaneous attraction and repulsion in the system (depicted by the shaded region in Fig. \ref{fig4}). This unusual dynamics arises due to a symmetry-breaking cooperative effect of the spiral waves. The relative motion induced by this effect is quite complicated. There are also some points, where $l$ is greater than half the critical distance, like for $l=10.2$ and $d=6.3$, where horizontal annihilation is observed (Fig. S1 in Supplementary section \cite{suppl}). These points are marked by leftward-pointing blue triangles of a smaller size. Although for this particular point, $l= 1.6 \times d$ and $l>\frac{d_c}{2}$, whereas $d<\frac{d_c}{2}$, still the spirals approach along $l$, and annihilate. It once again establishes the fact that spiral vortices feel a stronger attraction for their twins (born from the same initial wave), rather than other neighbors, which may be closer to them. However, the presence of the other spiral pair, enables this horizontal annihilation between the tips of the spiral wave, which would not have been achieved for a single spiral having $l>\frac{d_c}{2}$. Barring these few exceptions, for $l>\frac{d_c}{2}$ and $d>\frac{d_c}{2}$, there exists a large zone of spiral repulsion. With further increase in the vertical and horizontal distances (beyond $d_c$), all interaction vanishes. Here, from repulsive to no-interaction zone, the transition is very sharp with clear ($l$, $d$) demarcating lines, along the critical distance.

We have traced the trajectories of four numerical experiments, all lying in the repulsive zone from their initiation, till that point, where the relative motion of the tips cease [marked by circles in Fig. \ref{fig4}]. One may observe that the $l$ and $d$ values change spontaneously toward the zone of no-interaction. When both $l$ and $d$ values are much smaller than the critical value of 16.4 space units (e.g. the yellow circles which start from (11.55, 9.8), the vertical as well as the horizontal distance increases, as the pairs move toward the zone of no-interaction (16.4,16.4). They however do not trace an exact diagonal line. The vertical distance ($d$) increases faster than the horizontal ($l$), as seen from the curved path of the yellow circles. On the other hand, when one of the distances are beyond the critical distance, e.g. the experiment depicted by the green circles, where the initial distances are (19.4, 11.2), only the vertical distance increases all along the path until it reaches the steady state (19.4, 16.3).

Comparison of the phase diagrams of single and double spiral-pairs [Figs. \ref{fig1}(d) and \ref{fig4}] show some similarities and some dissimilarities. A careful observation of Fig. \ref{fig4} reveals that, for a fairly large value of $d$ ($> 16.4$), the system is in the no-interaction zone in the vertical direction. Again, for $l> 16.4$, it reaches the no-interaction zone in the horizontal direction. For both these conditions, the individual pair of spirals that lie close by, behave as they would in the absence of any other rotors [as in Fig. \ref{fig1}]. In the case of the former ($d>d_c$), as $l$ increases, we can observe horizontal annihilation followed by repulsion and then no-interaction. While in the latter case ($l>d_c$), vertical annihilation is followed by repulsion and no-interaction, as $d$ value is raised from small to large. The presence of the neighboring rotors for lower $d$ and $l$ values, brings about the complicated dynamics observed in the system [Figs. \ref{fig3}(b), \ref{fig3}(c) and S1] . From this observation, we can expect that an increase in the number of interacting spiral-cores might lead to more interesting dynamics.

\begin{figure}[t!]
\centering
\includegraphics[width=0.5\textwidth]{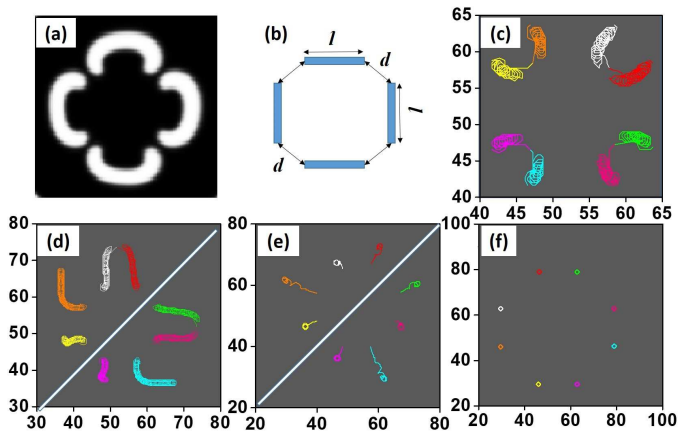}
\caption{Interactions between four pairs of spirals. (a) Snapshot of one numerical experiment at $t=40.8$ normalized time units, for $l=10.1$ and $d=6.9$ normalized space units. (b) System design showing the definition of $d$ and $l$ in these cases. (c) Tip trajectory for $l=10.1$ and $d=6.9$, (d) Tip trajectory for $l=9.0$ and $d=7.7$.  (e) Tip trajectory for $l=9.5$ and $d=12.5$ (for the purpose of simplicity, tip positions at intervals equal to the time period of the spirals, 5.34 time units, are only shown here). (f) Tip trajectory for $l=16.8$ and $d=23.2$.}
\label{fig5}
\end{figure}

In order to verify our hypothesis, we extended our system to eight spiral cores, or four spiral pairs. However, in order to avoid much complication, we designed a very symmetrical system of spirals. Figure \ref{fig5}(a), depicts such a system. In Fig. \ref{fig5}(b), we redefine our distance parameters $l$ and $d$ as the initial distance between the cores in a spiral pair, and shortest distance between two nearest cores belonging to different spiral pairs, respectively. We initiate four spiral pairs having the exact same dimension, and spaced equally apart, forming a kind of closed system, with a four-fold degeneracy in the initial wave forms [Fig. \ref{fig5}(a)].

In such a symmetric system, we observe mainly four kinds of interactions. A sample of each type has been illustrated in Fig. \ref{fig5}. Figure \ref{fig5}(c) shows mutual annihilation between diagonal pairs of spirals, due to a small $d$(= 6.9) value, as compared to $l$(= 10.1). In, Fig. \ref{fig5}(d),  $l$= 9.0 and $d$= 7.7, and we see both annihilation and repulsion due to a symmetry-breaking cooperative effect, similar to that seen in the case of two spiral pairs [\ref{fig3}(b)]. However, contrary to the two pair system, here we observe a mirror plane along the diagonal [white diagonal line in Fig. \ref{fig5}(d)]. It is noteworthy that the value of $d$ is smaller than $\frac{1}{2}d_c$, which would point towards an attractive interaction, in a simpler situation. Nonetheless, in addition to its own twin, the two tips of a spiral-pair is influenced by rotors belonging to two different spirals, placed at right angles on its either side. Figure \ref{fig5}(e) shows the repulsion of all the spirals. Here $l=9.5$ and $d=12.5$, both larger than $\frac{1}{2}d_c$. Although their direction of transition is quite symmetric, the velocity is different for different rotors. Some spirals have traveled larger distances compared to others, in an equal amount of time. Hence, the repulsion experienced by all the spirals are not uniform throughout. In this scenario also, we can observe the presence of a diagonal plane of symmetry. When the spirals are further away from each other, they do not show any visible  interaction, as depicted in Fig \ref{fig5}(f), where $l=16.8$ and $d=23.2$.

Preliminary studies with co-rotating spirals show similar trends in their interaction. However, they only demonstrate repulsive behavior, as spirals with the same sense of rotation are known not to attract each other. Even when they lie within one core length, such rotors repel each other, till they are separated by a critical distance, close to the $\lambda$ value.

\section{Experimental Methods}

The BZ reaction system provides a convenient way to study the behavior of spiral waves experimentally. A suitable concentration range that sustains spirals was chosen for our experiments. The final concentrations of the reactants are: [H$_2$SO$_4$] = 0.2 M, [NaBrO$_3$] = 0.04 M, [Malonic Acid] = 0.04 M and [Ferroin]= 0.001 M. We make a homogenous mixture of 0.8\% (w/v, final concentration) agarose gel in millipore water (having resistivity of 18.2 M$\Omega$~cm), with constant stirring and moderate heating. Then it is allowed to cool down just above the gelling  temperature with continued stirring so as to keep the mixture homogeneous throughout. Now, the other reactants (in water) are added to the stirred solution sequentially. The mixture is then poured into a Petri dish of 8 cm diameter and is allowed to cool. All experiments are carried out at room temperature (22$\pm$1 $^\circ$C). The reaction mixture is observed from above with a charge coupled device (CCD) camera (mvBlueFOX 22a), which is connected to a personal computer, while it is illuminated from below, with a white light source. We recorded the images at 2 second intervals which were later analyzed using in-house MATLAB scripts.

The recipe of the BZ reaction that was chosen for these experiments generated a non-meandering spiral wave with core length 0.09 cm (diameter), and an average wavelength of 0.48 cm. Single spiral pairs were generated in the usual way by cleaving a circular wave. In order to generate two pairs of spirals, at first two circular waves are initiated in close proximity of each other, by dipping two silver wires into the reaction gel for a few seconds. The silver helps in catalyzing the reaction, and hence initiates a circular target wave. The waves are allowed to expand and to come closer together. Finally, they are cleaved in such a way that we generate two pairs of spirals, facing one another. The distance between the circular waves at the time of cleaving has a special importance and we call it the distance of minimum approach. The value of $d$, or the vertical distance between the center of the spiral cores, depends on this distance of minimum approach. One more important parameter is the horizontal distance between the centers of the two spiral cores in a spiral pair generated from a single circular wave, ($l$). Special care was taken to maintain the symmetry in the system, by controlling the waveforms, so as to obtain similar $d$ and $l$ values between both pairs.

\begin{figure}[t!]
\centering
\includegraphics[width=0.45\textwidth]{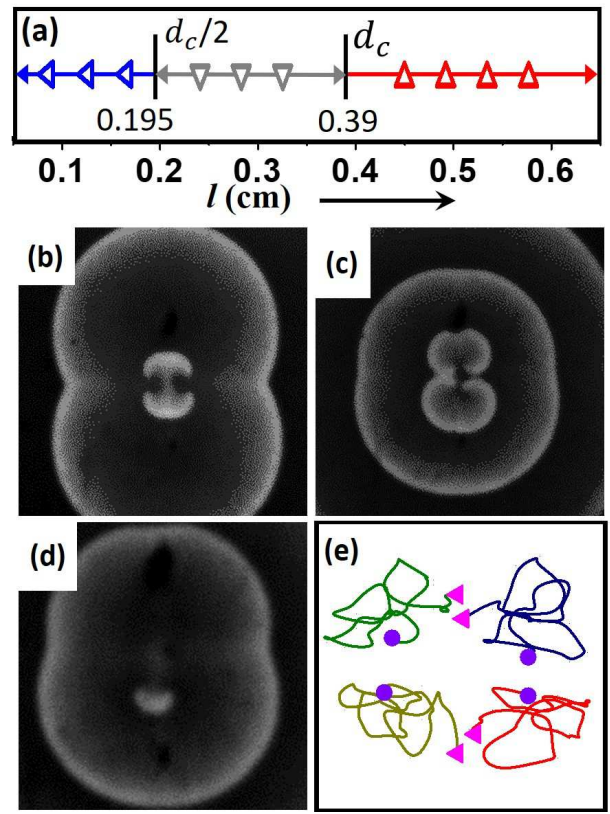}
\caption{(a) Phase diagram of interaction between the two cores of a spiral pair, in experiments. Blue left-pointing triangles depict annihilation, grey down-ward pointing triangles portray repulsion, and the open red up-ward pointing triangles represent the no interaction between the two rotors. (b)-(e) Annihilation of two pairs of spiral waves. Snapshots at (b) 8.47 min, (c) 27.28 min, and (d) 32.14 min after initiation of the reaction. Area of the snapshots are 2.95 cm $\times$ 2.95 cm. (e) Tip trajectories (colored curves) showing attraction and annihilation. The trajectory of each tip has been given a unique color for the purpose of clarity. Closed purple circles designate the initial position of every individual rotor, and the cyan triangles are the final positions prior to the moment of annihilation (at 27.33 min). Area shown in box is 0.55 cm $\times$ 0.55 cm. Initially, $d=0.14$ cm, $l=0.185$ cm. }
\label{fig6}
\end{figure}

\section{Experimental Results}

Our simulation results with two spirals [Fig. \ref{fig1}] was verified by our experiments with a single spiral wave having two counter-rotating spirals at its either end. We have designed a phase diagram for a pair of spirals in an experimental system [Fig. \ref{fig6}(a)], in keeping with the results of the numerical simulations, in order to graphically demonstrate the various kinds of interactions observed in our experiments. The critical distance, $d_c = 0.39$ cm, is one beyond which all interactions cease, and the distance $d_c/2$ marks the switch between repulsion and attraction. Analyzing with respect to the wavelength and core-size, as we did in our simulations, here the critical distance of interaction is equal to $d_c=\lambda-d_s= 0.48-0.09=0.39$ cm. So, the inter-core distance below which we should observe annihilation, is expected to be $\frac{1}{2} d_c=0.195$ cm. Here, we discuss in details, the systems with more number of rotors. We carried out a series of experiments with two pairs of spirals, by varying the distances $d$ and $l$,  that allowed us to observe the different kinds of interactive phenomena predicted by the simulations. We discuss representative examples of each kind here. However, the initial conditions were not always as symmetric as in the case of simulations. For example, the distance $l$($=0.185$ cm) in the experiment shown in Fig. \ref{fig6}, is an average of 0.17 cm (top pair) and 0.20 cm (bottom pair).

\begin{figure}[t!]
\centering
\includegraphics[width=0.45\textwidth]{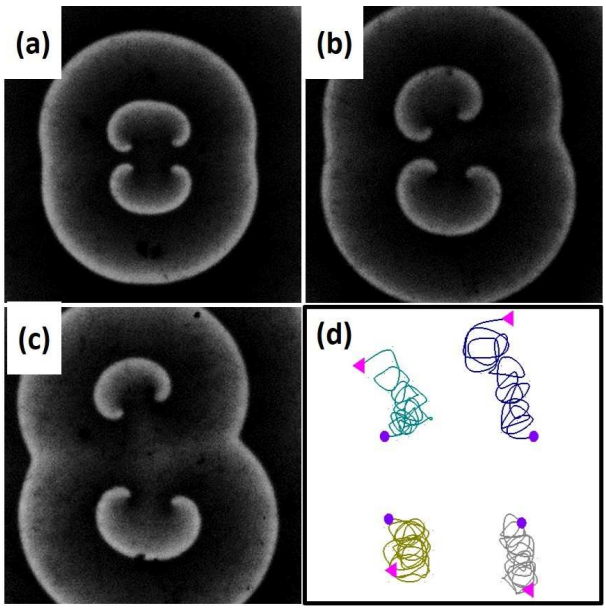}
\caption{Repulsion between two pairs of spiral waves. Snapshots at (a) 19 min, (b) 88 min, and (c) 174 min after initiation of the reaction. Area of the snapshots are 3.8 cm $\times$ 3.8 cm. (d) Tip trajectories showing repulsion. The circles and triangles designate the initial and later (at 72.0 min) positions of the individual rotors, respectively. Area shown in box is 0.75 cm $\times$ 0.75 cm. Initially, $d=0.295$ cm, $l=0.31$ cm.}
\label{fig7}
\end{figure}

Figures \ref{fig6}(b)-(d) depict the time evolution of two spiral pairs leading to annihilation. This is similar to the example shown in Fig. \ref{fig2}. The tip trajectories in Fig. \ref{fig6}(e) portray how a pair of rotors annihilate along the distance $l$. Here the initial distances are $d=0.14$ cm and $l=0.185$ cm. It is to be noted here, that attraction leading to annihilation occurs even when the initial distance between the rotors is more than two times the core length ($l= 2.05 \times $ core length). The minimum distance between the two spiral tips in the horizontal direction ($l_{min}$) are 0.12 cm (top pair) and 0.11 cm (bottom pair). While the minimum vertical distance in both the left and right pairs of tips ($d_{min}$) is 0.10 cm. All these distances are also larger than the core length. Even though the minimum distance in the vertical direction is close to a core length (0.09 cm), and smaller than the horizontal distance, the spiral tips attract along the horizontal, once again establishing the fact that the force of attraction is much stronger amongst those rotors that evolve out of a single circular wave. In this experiment however, both $l$ and $d$ are less than $\frac{1}{2}d_c$.

\begin{figure}[t!]
\centering
\includegraphics[width=0.45\textwidth]{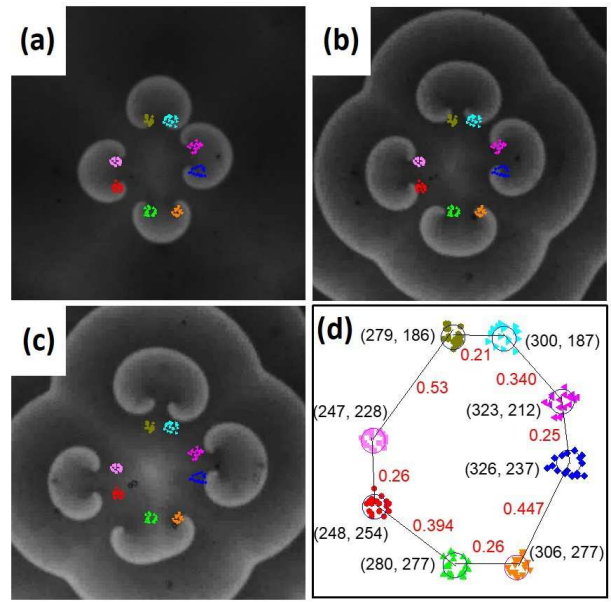}
\caption{Repulsion between four pairs of spiral waves. Snapshots covering an area of 2.95 cm $\times$ 2.95 cm, at (a) 6.07 min, (b) 56.57 min, and  (c) 138.93 min, after the initiation of the reaction. (d) Initial positions of the spiral cores (round curves tracing the dots, which are the positions of the spiral tip during the first rotation of the vortex). The coordinates of the center of the circular cores is been noted in pixels (in black), while the distance (in cm) between the center of the cores is given in red. The cores have also been juxtaposed over the snapshots.}
\label{fig8}
\end{figure}

On the other hand, the experiment illustrated in Fig. \ref{fig7} shows an increasing distance between spiral tips with time. This experiment is an example of spiral repulsion, as earlier seen in our simulations [Fig. \ref{fig3}(c)]. The values of $d$ and $l$  are 0.295 cm and 0.31 cm, respectively. Although $d$ and $l$ values are almost equal here, the tips repel each other vertically. Here, the horizontal minimum distance of approach are 0.22 cm (top pair) and 0.20 cm (bottom pair), while the minimum vertical distances between tips are 0.21 cm (left pair) and 0.22 cm (right pair). Due to a slight asymmetry in the initial conditions ($l=0.30$ cm in top pair, and 0.32 cm in bottom pair and $d$=0.29 cm in left pair and 0.30 cm in right pair), the symmetry is further broken in the system, as the reaction progresses. The tip trajectories [Fig. \ref{fig7}(d)] exhibit the divergent, yet un-symmetrical dynamics of the rotors over time. All initial distances here ($d$ and $l$) lie between $\frac{1}{2} d_c$ and $d_c$. Hence, the repulsion of the tips are in keeping with our theoretical predictions.

\begin{figure}[t!]
\centering
\includegraphics[width=0.45\textwidth]{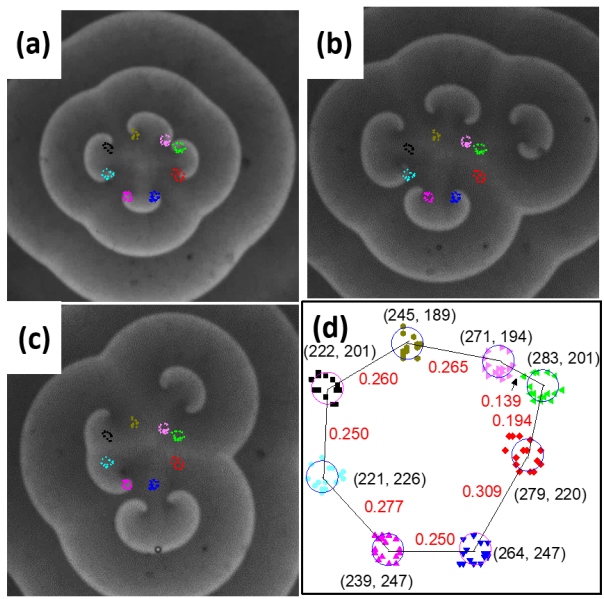}
\caption{ Annihilation of one pair of rotors following a strong repulsive interaction between four pairs of spiral waves. Snapshots at (a) 19 min (b) 116.1 min, and (c) 188.47 min after initiation of the reaction. Each snapshot covers an area of 2.95 cm $\times$ 2.95 cm. (d) Initial positions of the spiral cores  with the coordinates of the centers mentioned in pixels. The distance (in cm) between the center of each core with its two nearest neighbors is given in red. The initial cores of the eight rotors have also been superimposed over the snapshots.}
\label{fig9}
\end{figure}

We also carried out experiments with four spiral-pairs. We initiated the waves by cleaving four circular waves placed adjacent to each other, in a square arrangement. The first example is Fig. \ref{fig8}, where only repulsive interaction was observed between the spirals. The $d$ values here ranged between 0.34 cm to 0.53 cm, while the $l$ values ranged in the order of 0.21 cm to 0.26 cm. All $d$ values, except one (0.34 on the top right) are greater than the critical distance ($d_c=0.39$ cm), while the $l$ values are all less than $d_c$, but greater than $\frac{1}{2} d_c$. One would predict the spiral rotors to repel their twins (born from same wave), and not interact with the other neighbors. However, the dynamics gets complicated here. We observe that all the spiral rotors move away from the center, while (almost) maintaining the symmetry. As $l$ values increase very slightly between the tips, the $d$ values are seen to increase more. The only spiral pair that somewhat breaks this symmetry, is the one initiated at $y=277$, the right tip [initially at (306, 277)] of which is probably pinned to a small bubble that has been formed at its vicinity (seen just below the orange core of the tip in Fig. \ref{fig8}(c)]). This stops its expected movement in the downward direction like the other rotors, while the left tip travels downward, and the $l-$ value for this pair increases appreciably.

Figure \ref{fig9} shows an experiment where a strong repulsive interaction was observed, followed by annihilation of a wave-pair. The range of $d$ values in this example is 0.139 cm - 0.309 cm, while $l$ ranges between 0.194 cm and 0.265 cm. This system has an inherent dissymmetry right from the initiation [Fig. \ref{fig9}(a) and (d)]. Hence one might expect some very unique dynamics arising out of the wave-interactions. A quick glance at the initial inter-rotor distances point at only two values ($d = 0.139$ cm in the top-right corner and $l = 0.194$ for the right spiral pair) to be in the attractive range. All other distances are in the repulsive zone.

Both the attracting pairs have one rotor in common (core depicted as green, left-pointing triangles). So, there will be a competition for attracting this one rotor, by its two neighbors. As time progresses, the spiral wave-pair on the right is expelled to the further right, by the rest of the rotors [Fig. \ref{fig9}(b)]. This is an unexpected movement of the green rotor. Even though it was very close to its pink neighbor (core depicted by right-pointing triangles) [Figs. \ref{fig9}(a) and (d)], it chose to move away from it, and travel with its twin (the core of red diamonds). At this stage, the three spiral wave-pairs that remain at the center of the reaction chamber, display a three-fold symmetry among themselves [\ref{fig9}(b)]. Subsequently, the spiral wave at the top, eventually rotates (as a pair) in the clockwise direction, as it also moves away from the center. Meanwhile, the spiral-pair which had been expunged to the right, moves further away, and its two rotors start experiencing mutual attraction and the two vortices undergo annihilation at around 138.5 min (after the initiation of the reaction). This annihilation of the wave may be attributed to its $l$- value of 0.194 cm, which is marginally lower than $\frac{1}{2} d_c$. Please refer to the movie of the experiment in the Supplementary material to witness the very interesting wave dynamics of this experiment \cite{suppl}.

\section{Discussions and Conclusion}

We have carried out detailed analysis on the interaction of counter-rotating spiral pairs. The results we obtained have brought forth many new observations. Spiral waves, in close proximity, would attract each other, and finally annihilate. A slight change in initial distance could make two attracting rotors, highly repulsive. We have successfully established a critical distance of interaction between the rotor pairs, whether they have a single or multiple neighbors. While the transition from attractive to repulsive interaction is not so crisp for more than two rotors, for the simplest case of two rotors, this occurs at 8.2 space units in numerical simulation and 0.195 cm in experiments. As we increase the initial distance further, beyond a particular distance (16.4 space units and 0.39 cm), all interactions between the spirals cease to exist.

The results obtained for the interactive behavior of these two dimensional(2D) spiral waves, are somewhat different from that known about their three-dimensional(3D) counterparts, the scroll waves. In experiments with scroll rings, we had in an earlier study \cite{sdrecon}, shown that the circular filaments attract each other and reconnect, when they are less than one core length apart. Repulsion between the wave-forms was also found for a particular orientation of the filaments. However, it is difficult to establish quantitatively, the repulsive influence of the neighboring vortices on a shrinking scroll wave, as scroll rings having positive filament tension undergo spontaneous shrinkage, and eventually disappear. In another experimental study of parallel and straight scroll waves \cite{marcus3D}, it was shown that they repel each other when they are separated by distances greater than $\frac{2}{3} \lambda$ but smaller than one wavelength. Here, the authors could not initiate pairs of scroll waves that were closer than $\frac{2}{3} \lambda$, in their experiments. Hence, for 3D scroll waves, a critical distance for the sudden flipping of attractive interaction to give way to repulsion, is not yet known in literature.

Though attraction and repulsion among the 2D-spiral waves has also been observed in the current study, the main difference from the 3D system is that, here the attraction between rotors is felt over a distance that is many times more than the core length. Numerically the distance was found to be 4.5 times the core length, while experimentally we have observed attractive interaction atleast upto 2.2 times the core length. We have established that for 2D spiral rotors, the zone of attraction extends for a distance equal to half the critical distance, beyond which, repulsive interaction kicks in. This critical distance is equal to the difference of the wavelength and core size. In numerical simulations the value is 16.4 space units, and for our experiments, it was found to be 0.39 cm. When the distance between cores is more than this critical distance, there is no longer any interaction observed between the rotors.

We have also observed a priority of attractive interaction between the spiral rotors. Two spirals that are initially joined by a single wave, attract more strongly than they would, if they were not. It has been seen in numerical as well as experimental experiments, that a vortex is being attracted heavily by its twin, even though the distance separating it from a neighboring rotor (belonging to another spiral wave-pair) was much lower. This could be because of some underlying feedback occurring through the adjoining wave.

The symmetry-breaking dynamics of the spiral waves observed here are also extremely interesting. An initial dissymmetry, however small, blows up with time, and the system diverges into total asymmetry. The dominance of one spiral over the other in a spiral pair, leading to symmetry-breaking was earlier demonstrated in numerical simulations \cite{erma, aran}. However, with our experiments and simulations, we show that a system with multiple spirals, all having the same frequency can also undergo symmetry-breaking dynamics, even if the initial distances separating them are equal, leading to an uneven geometry as time progresses.

Spiral rotors of these kind play a vital role in the fibrillatory conduction of the atrial muscles, by activating the atria at exceedingly high frequencies \cite{jalife2}. The present study illuminates the nuances of spiral wave interaction, and may enable a better understanding of the interaction of rotors in the atria.

Further analysis of the velocity of attraction and repulsion should shed more light on the interaction dynamics. It remains to be seen, whether the velocity of these interactions would take the form of Yukawa potentials, as was found in the case of 3D scroll rings \cite{sdrecon}. Future studies could try to uncover the cause of the symmetry-breaking dynamics that is observed in systems with one, two and four spiral pairs.

\vspace{1 cm}
\section*{ACKNOWLEDGEMENTS}
This work was partially supported by the Science and Engineering Research Board, Government of India (Grant No. CRG/2019/001303).

\end{document}